\documentclass[prl,aps,twocolumn]{revtex4} 
\usepackage{graphicx} 
\usepackage{tabularx}
\usepackage{dcolumn} 
\usepackage{color}
\usepackage{bm}
\usepackage{amsmath}
\usepackage{amssymb}

\begin{document} 


\title{Notes on multiple superconducting phases in UTe$_2$  ---Third transition---
}

\author{Kazushige Machida} 
\affiliation{Department of Physics, Ritsumeikan University, 
Kusatsu 525-8577, Japan} 

\date{\today}

\begin{abstract}
A three-component Ginzburg-Landau theory for a triplet pairing is developed to understand the observed multiple phases in a new superconductor UTe$_2$ under pressure. Near the critical pressure $P_{\rm cr}$=0.2GPa where all components are perfectly degenerate the three successive superconducting transitions are predicted to occur. The $p$-wave pairing symmetry realized in UTe$_2$ is characterized by non-unitarity and chirality with point nodes, thus time reversal symmetry spontaneously broken.

 \end{abstract}

\maketitle 

Much attention has been focused on ferromagnetic superconductors (SC)~\cite{aoki-review}, UGe$_2$, URhGe, and UCoGe,
including a new spin-polarized heavy Fermion material UTe$_2$~\cite{ran,aoki2} 
which is believed to belong to the same category to the former three
compounds. A series of experiments on UTe$_2$ indicates that 
(1) a spin-triplet pairing is realized, (2) a pair of point nodes is orientated along the $a$-axis in orthorhombic
crystal (D$_{2h}$)~\cite{kittaka}, (3) it is chiral where the time reversal symmetry is spontaneously broken
and characterized by edge current. Thus the pairing symmetry is best described by the pair function $({\bf b}+i{\bf c})(p_b+ip_c)$
with double chiral non-unitary state both in the ${\bf d}$-vector of the spin space and in the orbital space.
This pairing function analogous to the superfluid $^3$He A$_1$ and A$_2$ phases explains several 
outstanding experimental facts, including  (1), (2) and (3) mentioned above.

Recently, Aoki, {\it et al}~\cite{aoki} and Braithwaite, {\it et al}~\cite{daniel} have discovered multiple phases in UTe$_2$ under pressure ($P$).
They enumerate four different SC phases in the $T$-$H$-$P$ space at least 
with $T$ and $H$ being temperature and applied field.
Since our previous theory~\cite{machida} can provide three phases, that is, the A$_1$, A$_2$ and A phases,
it needs to be extended to accommodate the additional phases.
This can be done quite naturally because our theoretical framework is 
based on the overall symmetry SO(3)$\times$D$_{2h}$$\times$U(1)
with the spin, orbital and gauge symmetry respectively.

This triple spin symmetry of the pairing function is expressed by a complex
three component vectorial order parameter $\vec{\eta}=(\eta_a,\eta_b,\eta_c)$.
Before~\cite{machida} we assumed the only two components are active, 
mainly aiming at explaining the $H_{\rm c2}$ phase diagrams, consisting of 
a reentrant SC in URhGe, an S-shape one in UCoGe, and an L-shape one in UTe$_2$.
Now the multiple phase diagram found requires clearly that all the three components should be active
afterall.

Under D$_{2h}$ symmetry the most general Ginzburg-Landau free energy functional up to the quadratic order is expressed
by

\begin{eqnarray}
F^{(2)}=\alpha_0(T-T_{\rm c0}){\vec \eta}\cdot{\vec \eta}^{\star}+b|{\vec M}\cdot{\vec \eta}|^2+
cM^2{\vec \eta}\cdot{\vec \eta}^{\star}\nonumber \\
+i\kappa {\vec M}\cdot {\vec \eta}\times {\vec \eta}^{\star}.
\nonumber
\label{e1}
\end{eqnarray}

\noindent
with $b$ and $c$ positive constants. We set $c$=0 without loss of generality. 
The last term arises due to the non-unitarity of the pairing function in the presence of the 
moment $\vec M$ which is to break the SO(3) spin symmetry. 
While the spontaneous ferromagnetic moment is established below the Curie temperatures
in UGe$_2$, URhGe, and UCoGe,
it is slowly fluctuating in UTe$_2$ which is detected by the NMR experiment by Tokunaga, et al~\cite{tokunaga}.
Upon lowering $T$ the fluctuations are wiped out via the $T_2$  measurement window, implying that the time scale of the
moment fluctuations is so slow, an order of MHz $\sim 10^{-5}$K. For the superconducting electrons with $T_{\rm c}=1.6$K
this time scale is slow enough to regard it as a symmetry breaker as explained before~\cite{machida}.

\begin{figure}
\includegraphics[width=6.4cm]{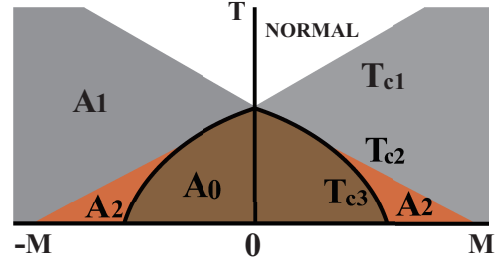}
\caption{(color online) 
Phase diagram in the $T$ and $M$ plane.
$T_{\rm c1}$ ($T_{\rm c2}$) for the A$_1$ (A$_2$) phase  increases (decreases) linearly in $M$.
The third phase A$_0$ decreases quadratically in $M$ away from the degenerate point at $M$=0.
}
\label{f1}
\end{figure}

It is convenient to introduce $\eta_{\pm}={1\over \sqrt2}(\eta_b\pm i\eta_c)$
for ${\bf M}=(M_a,0,0)$. From the above equation the quadratic term $F^{(2)}$ becomes

\begin{eqnarray}
F^{(2)}=\alpha_0\{(T-T_{\rm c1})|\eta_{+}|^2+(T-T_{\rm c2})|\eta_{-}|^2+(T-T_{\rm c3})|\eta_{a}|^2\}
\nonumber
\label{e2}
\end{eqnarray}

\noindent
with

\begin{eqnarray}
T_{\rm c 1,2}=T_{\rm c0} \pm{\kappa\over \alpha_0}M_a\nonumber\\
T_{\rm c 3}=T_{\rm c0} -{b\over \alpha_0}M^2_a
\nonumber
\label{e2}
\end{eqnarray}
\noindent
The actual second transition temperature may be modified because of the fourth order GL terms.
Here we note that among the GL fourth order terms, $Re(\eta_a^2\eta_{+}\eta_{-})$ becomes important
in interpreting the $H_{\rm c2}$ data later because it is independent of the signs of the GL parameters.

Figure~\ref{f1} shows that as a function of moment $M$, $T_{\rm c1}$ ($T_{\rm c2}$) increases (decreases) linearly,
while $T_{\rm c3}$ always decreases as $M$ grows. The three transition lines meet at $M$=0 where the 
three components $\eta_i$ are all degenerate. Thus away from the degenerate point at $M$=0, the A$_0$
phase starts at $T_{\rm c3}$ quickly disappears from the phase diagram.
Below $T_{\rm c2}$ the two components $\eta_{+}$ and $\eta_{-}$ coexist, symbolically denoted by
A$_1$+A$_2$. Note that because  their transition temperatures are different,
A$_1$+A$_2$ is not the so-called A-phase which is unitary, but generically non-unitary
except at the degenerate point $M$=0 where the totally symmetric phase is realized with time reversal symmetry preserved.
Likewise below $T_{\rm c3}$ all the components coexist; A$_1$+A$_2$+A$_0$ realizes.

\begin{figure}
\includegraphics[width=6.7cm]{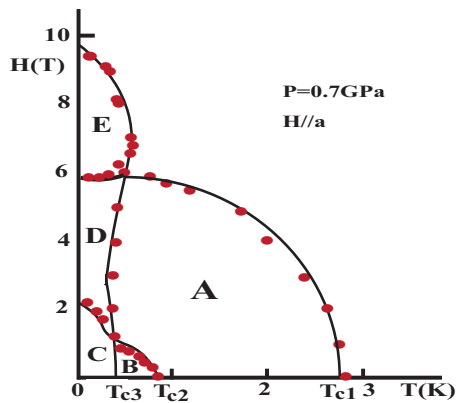}
\caption{(color online) Phase diagram in $H$-$T$ plane for $H\parallel a$-axis at $P$=0.7GPa.
The data points are from Ref.~[\onlinecite{aoki}].
In addition to $T_{\rm c1}$ and $T_{\rm c2}$ we identify $T_{\rm c3}$. Five phases are
enumerated.
}
\label{f2}
\end{figure}

Under an applied field with vector potential $\bf A$, the gradient GL energy is given 

\begin{eqnarray}
F_{grad}={\sum}_{\nu=a,b,c}\{K_a|D_x\eta_{\nu}|^2+K_b|D_y\eta_{\nu}|^2+K_c|D_z\eta_{\nu}|^2\}
\nonumber
\end{eqnarray}
\noindent
where $D_j=-i\hbar \partial_j-2eA_j/c$ and the mass terms are characterized by the coefficients $K_j$ ($j=a,b,c$) in D$_{2h}$.
It is seen from this form that $H_{\rm c2}$'s for the three components each starting at $T_{\rm c j}$ ($j=1,2, 0$) with different slopes
intersect each other, never avoiding or leading to a level repulsion.

We discuss the $H_{\rm c2}$ data of UTe$_2$ under pressure in light of this general rule.
As shown in Fig.~\ref{f2} where at $P$=0.7GPa for $H\parallel a$ the $H_{\rm c2}$ data points are quoted
from Ref.~[\onlinecite{aoki}] we draw the three continuous lines to connect those points,
finding the missing third transition along the $T$ axis.
Note that the tricritial point with three second order lines is thermodynamically forbidden~\cite{yip}.
We name each region A, B, C, D, and E. Those are characterized by A=A$_1$ at $T_{\rm c1}$, B=A$_1$+A$_2$ at $T_{\rm c 2}$,
C=A$_1$+A$_2$+A$_0$ at $T_{\rm c3}$, D=A$_1$+A$_0$, and E=A$_0$.
It is understood that this phase diagram is quite exhaustive, no further state expected
in our framework.
At the intersection points in Fig.~\ref{f2} the four transition lines should always meet 
together according to the above general rule and thermodynamics. The lines indicate how those three phases interact each other, by 
enhancing or suppressing. $T_{\rm c3}$ could be raised by the presence of A$_1$ and A$_2$ phases
due to the fourth order term $Re(\eta^2_a\eta_{+}\eta_{-})$ mentioned.

\begin{figure}
\includegraphics[width=6cm]{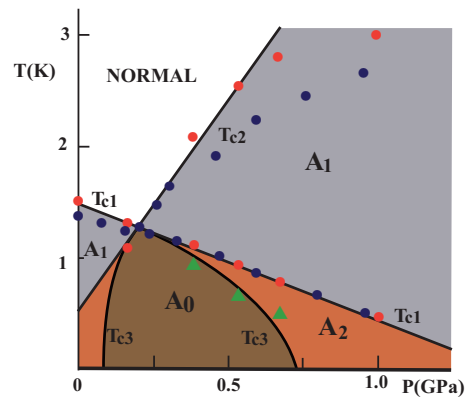}
\caption{(color online) Phase diagram in $T$-$P$ plane. The data points marked by red and blue dots
from Refs.~[\onlinecite{aoki}] and [\onlinecite{daniel}] respectively. The triangle symbols denote $T_{\rm c3}$ determined as
explained in the main text.
}
\label{f3}
\end{figure}

Figure 3 shows the phase diagram in the $P$-$T$ plane where $T_{\rm c3}$  determined 
thus for $P$=0.40, 0.54 and 0.70GPa is displayed by
triangle symbols. The identification for each phase  is done reasonably. The critical pressure 
$P_{\rm cr}$=0.2GPa where all phase transition lines meet is identified.
Away from $P_{\rm cr}$ the three transition lines split linearly for $T_{\rm c1}$ and $T_{\rm c2}$,
and quadratically for $T_{\rm c3}$. Note that at ambient pressure the second transition $T_{\rm c2}$
is probed by NMR experiment~\cite{nakamine}.

In conclusion, we have identified the third SC phase A$_0$ in  addition to  A$_1$ and A$_2$ 
through the GL analysis of the $H_{\rm c2}$
measurements under pressure. The chiral-$p$-wave pairing with non-unitary triplet state is found to be realized
in UTe$_2$ analogous to superfluid $^3$He A-phase under fields, thus time reversal symmetry is spontaneously broken.

{\bf Acknowledgments}
The author thanks D. Aoki for sharing experimental data prior to publication.
This work was supported by JSPS KAKENHI No.17K05553.

\end{document}